\input ./style/arxiv-general.cfg
\documentclass[aoas,MSNbibl,nameyear,dvips]{arximspdf}
\makeatletter
   \@ifpackageloaded{graphicx}{}{\usepackage{graphicx}}
\makeatother
\usepackage{dcolumn}

\doi{10.1214/15-AOAS829}
\volume{9}
\issue{2}
\pubyear{2015}
\firstpage{1076}
\lastpage{1101}
\docsubty{FLA}

\makeatletter
\newcolumntype{d}[1]{D{.}{.}{#1}}
\newcommand{\rrVert}{\Vert}
\newcommand{\llVert}{\Vert}
\newcommand{\eqref}[1]{(\ref{#1})}
\newcommand{\bs}{\mathbf{s}}
\newcommand{\by}{\mathbf{y}}
\newcommand{\bD}{\mathbf{D}}
\newcommand{\bI}{\mathbf{I}}
\newcommand{\bP}{\mathbf{P}}
\newcommand{\bR}{\mathbf{R}}
\newcommand{\bT}{\mathbf{T}}
\newcommand{\bU}{\mathbf{U}}
\newcommand{\bV}{\mathbf{V}}
\newcommand{\bX}{\mathbf{X}}
\newcommand{\bzero}{\mathbf{0}}
\newcommand{\bone}{\mathbf{1}}
\newcommand{\bbeta}{\bolds{\beta}}
\newcommand{\bdelta}{\bolds{\delta}}
\newcommand{\bgamma}{\bolds{\gamma}}
\newcommand{\bvarepsilon}{\bolds{\varepsilon}}
\newcommand{\brho}{\bolds{\rho}}
\newcommand{\bPi}{\bolds{\Pi}}

\newcommand{\argmin}{\arg\min}
\newcommand{\cov}{\operatorname{Cov}}
\makeatother

\begin{document}
\begin{frontmatter}

\title{Wavelet-domain regression and predictive inference in psychiatric neuroimaging}
\runtitle{Wavelet-domain image regression}

\begin{aug}
\author[A]{\fnms{Philip T.}~\snm{Reiss}\corref{}\thanksref{T1,M1,M2}\ead[label=e1]{phil.reiss@nyumc.org}},
\author[A]{\fnms{Lan}~\snm{Huo}\thanksref{T2,M1}\ead[label=e2]{huolanlan@gmail.com}},
\author[A]{\fnms{Yihong}~\snm{Zhao}\thanksref{M1}\ead[label=e4]{yihong.zhao@nyumc.org}},
\author[A]{\fnms{Clare}~\snm{Kelly}\thanksref{T2,M1}\ead[label=e5]{amclarekelly@gmail.com}}
\and
\author[B]{\fnms{R. Todd}~\snm{Ogden}\thanksref{T3,M3}\ead[label=e3]{to166@cumc.columbia.edu}}
\runauthor{P.~T. Reiss et al.}
\affiliation{New York University\thanksmark{M1}, Nathan S. Kline
Institute for Psychiatric Research\thanksmark{M2} and~Columbia
University\thanksmark{M3}}
\address[A]{P.~T. Reiss\\
L.~Huo\\
Y. Zhao\\
C. Kelly\\
Department of Child and Adolescent Psychiatry\\
New York University School of Medicine\\
1 Park Ave., 7th floor\\
New York, New York 10016\\
USA\\
\printead{e1}\\
\phantom{E-mail: }\printead*{e2}\\
\phantom{E-mail: }\printead*{e4}\\
\phantom{E-mail: }\printead*{e5}}
\address[B]{R.~T. Ogden\\
Department of Biostatistics\\
Columbia University\\
722 W. 168th St., 6th floor\\
New York, New York 10032\\
USA\\
\printead{e3}}
\end{aug}
\thankstext{T1}{Supported in part by NSF Grant DMS-09-07017 and
National Institutes of Health Grants 5R01EB009744-03 and 1R01MH095836-01A1.}
\thankstext{T2}{Supported in part by National Institutes of Health
Grant 1R01MH095836-01A1.}
\thankstext{T3}{Supported in part by National Institutes of Health
Grant 5R01EB009744-03.}

%
\received{\smonth{8} \syear{2013}}
%
\revised{\smonth{2} \syear{2015}}

%
\begin{abstract}
An increasingly important goal of psychiatry is the use of brain
imaging data to develop predictive models. Here we present two
contributions to statistical methodology for this purpose. First, we
propose and compare a set of wavelet-domain procedures for fitting
generalized linear models with scalar responses and image predictors:
sparse variants of principal component regression and of partial least
squares, and the elastic net. Second, we consider assessing the
contribution of image predictors over and above available scalar
predictors, in particular, via permutation tests and an extension of
the idea of confounding to the case of functional or image predictors.
Using the proposed methods, we assess whether maps of a spontaneous
brain activity measure, derived from functional magnetic resonance
imaging, can meaningfully predict presence or absence of attention
deficit/hyperactivity disorder (ADHD). Our results shed light on the
role of confounding in the surprising outcome of the recent ADHD-200
Global Competition, which challenged researchers to develop algorithms
for automated image-based diagnosis of the disorder.
\end{abstract}

%
\end{frontmatter}

\section{Introduction}\label{sec1}
A major goal of current psychiatric neuroimaging research is to predict
clinical outcomes on the basis of quantitative image data. Many studies
have focused on ``predicting'' current disease states from brain images
[e.g., \citet{craddock2009,sun2009}]. While seemingly less
difficult than accurate prediction of \emph{future} outcomes, the goal
of clinically useful imaging-based diagnosis has proved highly
challenging [\citet{kapur2012,honorio2012}].

This paper addresses two important limitations of standard methods for
using brain images to predict psychiatric outcomes:
\begin{longlist}[(ii)]
\item[(i)] Ordinarily, the voxels (volume units) of the brain are treated as
interchangeable predictors or ``features.'' Accuracy might be improved
by properly exploiting the spatial arrangement of the brain.

\item[(ii)] In some cases brain images may prove successful for diagnostic
classification, but only because the images are related to one or more
scalar covariates that drive the association. This is a nonstandard
form of confounding, and there seems to be no existing methodology for
detecting it. In other words, little is known about how to assess
whether image data offers ``added value'' for prediction, beyond what
is available from nonimage data---which will typically be much simpler
to acquire.
\end{longlist}

To address limitation (i), we approach the general problem as one of
regressing scalar responses on \emph{image predictors}, which are
viewed, as in \citet{reiss2006} and \citet{reiss2010}, as a
challenging special case of functional predictors [\citet
{ramsay2005}]. The responses $y_1,\ldots,y_n$ are assumed to be
generated independently by the model
%
\begin{eqnarray}
y_i &\sim& \mathit{EF}(\mu_i, \phi),\label{efspec}
\\
g(\mu_i)&=&\mathbf{t}_i^T\bdelta+ \int
_{\mathcal
S}x_i(\bs)\beta(\bs)\,d\bs.\label{gmuspec}
\end{eqnarray}
Here $\mathit{EF}(\mu_i, \phi)$ denotes an exponential family distribution
with mean $\mu_i$ and scale parameter $\phi$, along with a link
function $g$; $\mathbf{t}_i$ is an $m$-dimensional vector of
(scalar) covariates, of which the first is the constant 1;
$x_i\dvtx {\mathcal S}\longrightarrow{\mathbb R}$ is a functional predictor
with domain ${\mathcal S}\subset{\mathbb R}^2$ or $\subset{\mathbb
R}^3$; and the corresponding effect, the \emph{coefficient function}
or \emph{coefficient image} $\beta\dvtx {\mathcal S}\longrightarrow
{\mathbb R}$, is the parameter of interest.
The simplest special case is the linear model
%
\begin{equation}
y_i =\mathbf{t}_i^T\bdelta+ \int
_{\mathcal
S}x_i(\bs)\beta(\bs)\,d\bs+\varepsilon_i,\label{FLM}
\end{equation}
where the $\varepsilon_i$ are independent and identically distributed
errors with mean 0 and variance $\sigma^2(=\phi)$. When $\mathbf{t}_i\equiv1$ (i.e., no scalar covariates), model \eqref
{FLM} is the extension, from one-dimensional to multidimensional
predictors, of the functional linear model that has been studied by
\citet{marx1999}, \citet{cardot1999}, \citet
{muller2005}, \citet{ramsay2005}, \citet{hall2007},
\citet{reiss2007}, \citet{goldsmith2011} and many others.

For the case of one-dimensional functional predictors, a popular way to
take spatial information into account is to restrict $\beta(\cdot)$
to the span of a spline basis [e.g., \citet{marx1999}]. Spline
methods for two-dimensional predictors have been studied by \citet
{Marx2005} and \citet{Guillas2010}, and by \citet
{reiss2010}, whose work was motivated by neuroimaging applications.

Some more recent work has considered neuroimaging applications with
two- and three-dimensional predictors [\citet
{zhou2013,goldsmith2013,huang2013}]. In this paper, we propose a set of
new approaches based on a wavelet representation of the coefficient
image. The idea of transforming the images to the wavelet domain
has previously appeared in the brain mapping literature, where it is
customary to fit separate models at each voxel, with the image-derived
quantity regressed on demographic or clinical variables of interest
[e.g., \citet{ruttimann1998,vandeville2007}]. But for our
objective of using entire images in a \emph{single} model to predict a
scalar response, working in the wavelet domain has been mentioned as a
natural idea [\citet{grosenick2013}] but rarely if ever pursued,
at least until the very recent work of \citet{wang2014}.
Unlike spline bases, wavelet bases are designed for sparse
representation and yield estimates that adapt to the features of the
coefficient image.

Limitation (ii) was highlighted by the results of the recent ADHD-200
Global Competition for automated diagnosis of attention
deficit/hyperactiv\-ity disorder [\citet{adhd2012}]. Teams were
provided with functional magnetic resonance images from ADHD subjects
and controls on which to train diagnostic algorithms, and then applied
these algorithms to predict diagnosis in a separate set of images. A
team of biostatisticians from Johns Hopkins University, whose methods
are described by \citet{eloyan2012}, achieved the highest score
for correct imaging-based classification and were declared the winners.
But a team from the University of Alberta, which discarded the images
and used just four scalar predictors [age, sex, handedness and IQ; see
\citet{brown2012}], attained a slightly higher classification
score [see \citet{caffo2012} for related discussion].

To address limitation (ii), we test the effect of image predictors via
a permuta\-tion-based approach originally proposed in the neuroimaging
literature [\citet{golland2003}], which we extend to allow for
scalar covariates. We also consider how to extend the traditional
notion of confounding to settings with both scalar and image
predictors. These ideas are illustrated using our wavelet methods, but
are not specific to them; rather, they are applicable with other
approaches to functional or high-dimensional regression.

Our contributions can be summarized as follows: (i) We propose novel
wavelet-domain methodology for regression with image predictors. While
\citet{wang2014} and \citet{zhao2014} have studied the
wavelet-domain lasso for image predictors, we also propose and compare
several other methods, and consider the generalized linear case and the
role of scalar covariates. (ii) We extend predictive performance-based
hypothesis testing [\citet{golland2003}] to the case where scalar
confounders are present, providing a new way to assess the usefulness
of image-based prediction.

In Section~\ref{wave} we introduce wavelet bases, and motivate and
outline a general template for scalar-on-image regression in the
wavelet domain. Section~\ref{twda} describes three specific
algorithms, which are evaluated in simulations in Section~\ref
{simsec}. Section~\ref{infsec} considers hypothesis testing and
confounding with image predictors. In Section~\ref{appsec} the
proposed methods are applied to a portion of the ADHD-200 data set, and
the results point to a possible role of confounding in the
competition's surprising result. Section~\ref{discuss} offers a
concluding discussion.

\section{Wavelets and their use in regression on images}\label{wave}
\subsection{A brief introduction to wavelet basis representations}
\begin{figure}[b]

\includegraphics{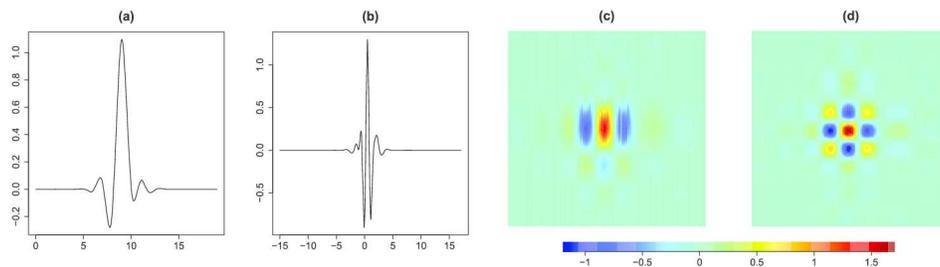}

\caption{\textup{(a)} Scaling function $\phi$ and \textup{(b)} wavelet function $\psi$
for 1D \citet{daubechies1988} ``least-asymmetric'' wavelets with
10 vanishing moments. 2D basis functions are formed from tensor
products such as \textup{(c)} $(x,y)\mapsto\psi(x) \phi(y)$ and \textup{(d)}
$(x,y)\mapsto\psi(x) \psi(y)$.}
\label{figdaub}
\end{figure}

Wavelet bases are a popular way to obtain a sparse representation for
functional data, in particular, when the degree of smoothness exhibits
local variation [see \citet{ogden1997,vidakovic1999,nason2008}
for statistically-oriented treatments]. A wavelet basis for
$L^2(\mathbb R)$ is constructed from a scaling function (or ``father
wavelet'') $\phi$ and a wavelet function (``mother wavelet'') $\psi$
[see Figure~\ref{figdaub}(a)--(b)], with the following properties:
\begin{itemize}
\item
For each $j\in\mathbb Z$, the shifted and dilated functions $\{\phi
_{j,k}(x) = 2^{j/2} \phi(2^jx - k)\dvtx k\in{\mathbb Z}\}$ form an
orthonormal basis for $V_j$, where $\cdots\subset V_{-1}\subset V_0
\subset V_1\subset\cdots$ is a nested sequence of subspaces whose
union is a dense subspace of $L^2(\mathbb R)$.
\item For each $j\in\mathbb Z$, $\{\psi_{j,k}(x) = 2^{j/2} \psi(2^jx
- k)\dvtx k\in{\mathbb Z}\}$ form an orthonormal basis for a ``detail
space'' $W_j$ satisfying $V_{j+1}=V_j\oplus W_j$.
\end{itemize}
Hence, for any integer $j_0 \ge0$, $V_{j_0}\oplus W_{j_0}\oplus
W_{j_0+1}\oplus\cdots$ is a dense subspace of $L^2(\mathbb R)$.
%

Given appropriate boundary handling, such as modifying the scaling and
wavelet functions to be periodic, one can likewise construct
orthonormal wavelet bases for $L^2[0,1]$, of the form
\[
\{\underbrace{ \phi_{j_{0},0}, \ldots, \phi_{j_0,2^{j_0}-1}}_{\in
V_{j_0}}
\}\cup\{\underbrace{\psi_{j_0,0},\ldots, \psi_{j_0,
2^{j_0}-1}}_{\in W_{j_0}}
\}\cup\{\underbrace{\psi_{j_0+1,0}, \ldots, \psi_{j_0+1,2^{j_0+1}-1}}_{\in W_{j_0+1}}
\} \cup\cdots %
\]
---that is, $2^{j_0}$ scaling functions (corresponding to the
large-scale features of the data), $2^{j_0}$ wavelet functions at level
$j_0$, $2^{j_0+1}$ wavelet functions at level $j_0+1$ and so on, with
higher wavelet levels capturing finer-scale details. This multiscale
structure is what makes wavelet bases so useful for sparse
representation of functions with varying degrees of smoothness.

The wavelet decomposition level $j_0$ acts as a tuning parameter. A
small $j_0$ implies that a small number ($2^{j_0}$) of scaling
functions are used to construct the macro features of the function,
with most of the basis elements dedicated to providing detail at a
variety of scales. A large $j_0$ allows for many more scaling
functions, each at higher resolution, and thus fewer basis elements
corresponding to detail.

In practice, a function $f\in L^2[0,1]$ is observed at finitely many
points, ordinarily taken to be the $N=2^J$ (for some positive integer
$J$) equally spaced points $0,\frac{1}{N},\ldots,\frac{N-1}{N}$.
(When the function is observed at a number of points that is not a
power of 2, one can insert zeroes before and after to attain the next
highest power of~2.) The observed values can then be interpolated by
the $N$-dimensional truncated basis
\[
\{ \phi_{j_{0},0}, \ldots, \phi_{j_0,2^{j_0}-1}\}\cup\{\psi _{j_0,0},
\ldots, \psi_{j_0, 2^{j_0}-1}\}\cup\cdots\cup\{\psi _{J-1,0}, \ldots,
\psi_{J-1,2^{J-1}-1}\}.
\]
The discrete wavelet transform (DWT), implemented by the $O(N)$ pyramid
algorithm of \citet{mallat1989}, expands $f$ with respect to this
basis. Given a judicious choice of $\phi$ and $\psi$, signals of
varying smoothness can be well represented with a small number of
coefficients. Throughout this paper we use the compactly supported
\citet{daubechies1988} ``least-asymmetric'' wavelets with 10
vanishing moments, displayed in Figure~\ref{figdaub}.

Wavelet bases for two dimensions can be constructed by taking tensor
products of the $\phi$ and $\psi$ functions. The two-dimensional
scaling function is $\phi(x) \phi(y)$ and there are three types of 2D
wavelets: $\phi(x) \psi(y)$, $\psi(x) \phi(y)$ and $\psi(x) \psi
(y)$, roughly corresponding to ``horizontal,'' ``vertical'' and
``diagonal'' detail, respectively [see Figure~\ref{figdaub}(c)--(d)].
These functions are dilated and translated just as their 1D
counterparts are. Wavelet bases for 3D are constructed similarly.
\citet{morris2011} discuss alternative wavelet transforms for
images that are not constructed as tensor products.

\subsection{A meta-algorithm for scalar-on-image regression}\label{metalg}
Henceforth, the functional predictor $x_i(\cdot)$ of \eqref{gmuspec},
\eqref{FLM} will be replaced by the $i$th discretized image
observation $\mathbf{x}_i=(x_1,\ldots,x_N)^T\equiv
[x_i(s_1),\ldots,x_i(s_N)]^T$, where $s_1,\ldots,s_N\in\mathcal S$
are distinct spatial locations at which the function $x_i$ is measured.
Often, in practice, each image is given as a matrix or 3D array; $\mathbf{x}_i$ is then obtained by converting this into a vector.
From now until Section~\ref{x2glm} we focus on the linear model \eqref
{FLM}, which can now be written in matrix form as
%
\begin{equation}
\label{matrixform} \by= \bT\bdelta+ \bX\bbeta+ \bvarepsilon.
\end{equation}
Here $\by= (y_1, \ldots, y_n)^T$; $\bvarepsilon=(\varepsilon_1,
\ldots, \varepsilon_n)^T$; $\bT$ is the $n \times m$ matrix with
$i$th row $\mathbf{t}_i^T$; $\bX$ is the $n \times N$
matrix with $i$th row $\mathbf{x}_i^T$; and $\bbeta=(\beta
_1, \ldots, \beta_N)^T$ is a similarly discretized version of the
coefficient image $\beta$. More\vspace*{1pt} precisely, for $j=1,\ldots,N$, $\beta
_j=w_j\beta(s_j)$, where the $w_j$'s are quadrature weights such that
$\mathbf{x}_i^T\bbeta$ is a good approximation to the
integral in \eqref{FLM}; but for image data, $s_1,\ldots,s_N$
typically form an equally spaced grid, so these weights are taken as
constant and hence ignored. With these definitions, \eqref{FLM} is
just the $i$th of the $n$ equations that make up the vector equation
\eqref{matrixform}.

To simplify the notation, we shall use a single subscript and denote
the wavelet basis functions for a given $j_0$ as $\{ \psi_1, \psi_2,
\ldots, \psi_N \}$. The wavelet representation of the $i$th observed
image $x_i$ is $x_i(s)=\sum_{k=1}^N \tilde{x}_{ik} \psi_k(s)$, in
which the
\emph{wavelet coefficients} are given by
$\tilde{x}_{ik} = \langle x_i, \psi_k \rangle$. The coefficient
vector $\tilde{\mathbf{x}}_i=(\tilde{x}_{i1},\ldots,\tilde{x}_{iN})^T$ can be written as $\tilde{\mathbf{x}}_i = {\mathcal W} \mathbf{x}_i$,
where ${\mathcal W}$ is an $N \times N$ orthonormal matrix (which is
not formed explicitly when $\tilde{\mathbf{x}}_i$ is
computed by the DWT).
Similarly the discretized coefficient function $\bbeta$ can be
represented in terms of its wavelet coefficients as $\tilde{\bbeta} =
{\mathcal W} \bbeta$, leading to the wavelet-domain version of model
\eqref{matrixform}:
%
\begin{eqnarray}\label{wdform}
\nonumber
\by&=& \bT\bdelta+ \bX{\mathcal W}^T{\mathcal W}\bbeta+
\bvarepsilon
\nonumber\\[-8pt]\\[-8pt]\nonumber
&=&\bT\bdelta+ \tilde{\bX}\tilde{\bbeta} + \bvarepsilon,
\end{eqnarray}
where $\tilde{\bX}$ is the $n\times N$ matrix with $i$th row $\tilde
{\mathbf{x}}_i^T$.

The key point is that the wavelet-domain form \eqref{wdform} is better
suited than the original form \eqref{matrixform} for applying sparse
techniques for high-dimensional regression---both because wavelet bases
are designed for sparse representation of images [\citet
{mallat2009}] and because the DWT approximately decorrelates or
``whitens'' data [\citet{vidakovic1999}]. We can thus formulate a
``meta-algorithm'' for scalar-on-image regression in the wavelet domain:
\begin{longlist}[3.]
\item[1.] Apply the DWT to the image predictors to transform model \eqref
{matrixform} into model~\eqref{wdform}.

\item[2.] Use some high-dimensional regression methodology to derive a
sparse estimate~$\hat{\tilde{\bbeta}}$.

\item[3.] Apply the inverse DWT to $\hat{\tilde{\bbeta}}$ to obtain a
coefficient image estimate $\hat{\bbeta}$ for the original model
\eqref{matrixform}.
\end{longlist}
Different choices for step 2 lead to specific algorithms, as described
in the next section.

The above general scheme can be extended to multiple image predictors
[cf. \citet{zhu2010}]. We note that this meta-algorithm has been
applied before for 1D functional predictors [\citet
{brown2001,wang2007,malloy2010,zhao2012}] and more for image predictors
[\citet{wang2014,zhao2014}]. Past work on wavelet-domain
classification, as opposed to regression [e.g., \citet
{berlinet2008,zhu2012,chang2014}], may bear comparison to our proposed
methods. \citet{morris2011} develop wavelet-domain functional
mixed models with images as \emph{responses}.

\section{Three wavelet-domain algorithms}\label{twda}
\subsection{Sparse wavelet-domain principal component regression}
The functional linear model \eqref{FLM} is often fitted by assuming
the coefficient function has a truncated functional principal
component, or Karhunen--Lo\`{e}ve, representation $\beta(s)=\sum_{j=1}^m c_j\rho_j(s)$, where $m$ is a positive integer and $\rho
_1,\rho_2,\ldots,\rho_m$ are the first $m$ eigenfunctions of the
covariance operator associated with the predictor functions $x_i$
[e.g., \citet{cardot1999,muller2005,cai2006}]. The eigenfunctions
$\rho_1,\rho_2,\ldots,\rho_m$ can be estimated by viewing the
functional predictors as (highly) multivariate data, and applying
ordinary principal component analysis to the predictor matrix $\bX$.

Here and in Section~\ref{swdpls}, we assume that $\bX$ has
mean-centered columns, that is,
$\bone^T\bX=\bzero$.
The approach of the previous paragraph then amounts to assuming $\bbeta
=\bV_m\bgamma$ for some $\bgamma\in{\mathbb R}^m$, where $\bU\bD
\bV^T$ is the singular value decomposition of $\bX$, and $\bV_m$
comprises the leading $m$ columns of $\bV$. Hence, estimation reduces
to choosing $\bdelta,\bgamma$ to minimize the principal component
regression [PCR; \citet{massy1965}] criterion
%
\begin{equation}
\label{pcrcrit}\llVert \by- \bT\bdelta- \bX\bV_m\bgamma \rrVert
^2.
\end{equation}
(This is a slightly nonstandard PCR criterion, in that principal
component reduction is applied only to $\bX$ but not to $\bT$. A
similar remark applies to the other criteria introduced below.)

As shown by \citet{reiss2007}, PCR can be implemented more
effectively by exploiting the functional character of the data. In the
one-dimensional functional predictor case, this has usually meant
forming smooth estimates of the eigenfunctions---as in the FPCR$_C$
method of \citet{reiss2007}, which expands the eigenfunctions
with respect to a $B$-spline basis [cf. \citet{cardot2003}]. But
for image predictors, local adaptivity---the ability to capture sharp
features in some areas vs. a high degree of smoothness
elsewhere---becomes particularly important. This motivates using a
wavelet basis, rather than a spline basis, to represent the
eigenfunctions, or, in other words, developing a wavelet-domain version
of PCR as an instance of the meta-algorithm of Section~\ref{metalg}.

A \emph{non}sparse wavelet-domain PCR estimate would minimize
%
\begin{equation}
\label{nspcrcrit}\llVert \by- \bT\bdelta- \tilde{\bX} \tilde{\bV}_m
\bgamma\rrVert ^2,
\end{equation}
which is analogous to \eqref{pcrcrit} but based on the SVD of $\tilde
{\bX}$ rather than of $\bX$. However, the advantage of working in the
wavelet domain is to obtain a sparse coefficient estimate by replacing
the PC weights $\tilde{\bV}_m$ with weights from a sparse version of
PCA. Several penalty-based methods have been proposed for sparse PCA
[e.g., \citet{zou2006,shen2008,witten2009}], but we opted for the
approach of \citet{johnstone2009}, which is simpler than the
penalized methods and, unlike them, was developed with a view toward
sparse wavelet representations of signals. \citet{johnstone2009}
propose to select the features or coordinates with highest variance,
and apply PCA only to these. The resulting sparse PCR criterion is
%
\begin{equation}
\label{spcrcrit}\bigl\llVert \by-\bT\bdelta- \tilde{\bX}^* \tilde{
\bV}^*_m\bgamma\bigr\rrVert ^2;
\end{equation}
here $\tilde{\bX}^*$ consists of the $c$ columns of $\tilde{\bX}$
having highest variance, and $\tilde{\bV}^*_m$ consists of the
leading $m$ columns of $\tilde{\bV}^*$, where $\tilde{\bU}^*\tilde
{\bD}^*\tilde{\bV}^{*T}$ is the SVD of $\tilde{\bX}^*$. The
minimizer $(\hat{\bdelta},\hat{\bgamma})$ of \eqref{spcrcrit} can
be obtained by simple least squares. The vector of wavelet coefficient
estimates is then $\hat{\tilde{\bbeta}}=\tilde{\bV}^*_m\hat
{\bgamma}$, and the coefficient image estimate $\hat{\bbeta
}={\mathcal W}^T\hat{\tilde{\bbeta}}$ is derived by the inverse DWT.

\subsection{Sparse wavelet-domain partial least squares}\label{swdpls}
Whereas PCR reduces dimension by regressing on the leading PCs of the
predictors, partial least squares [PLS; \citet{wold1966}] works
by regressing on a set of components that are relevant to predicting
the responses. A (nonsparse) wavelet-domain PLS estimate [cf.
\citet{nadler2005}] is derived by minimizing
%
\begin{equation}
\label{nsplscrit}\llVert \by-\bT\bdelta-\tilde{\bX} \tilde{\bR}_m\bgamma
\rrVert ^2
\end{equation}
[cf. \eqref{nspcrcrit}], where the columns of $\tilde{\bR}_m$ are
defined iteratively as follows [\citet{stone1990}]:
\begin{itemize}
\item$\tilde{\mathbf{r}}_1=\argmin_{\llVert  r\rrVert  =1}\cov(\by,\tilde{\bX}\mathbf{r})$;
\item for $j=2,\ldots,c$,
\[
\tilde{\mathbf{r}}_j=\argmin_{\llVert  r\rrVert  =1,\mbox{ }r^T\tilde
{X}^T\tilde{X}r_m=0 ~\forall m=1,\ldots,j-1}\cov(\by,\tilde {\bX}
\mathbf{r}).
\]
\end{itemize}

Once again, however, the point of working in the wavelet domain is to
obtain a sparse estimate. To define sparse wavelet-domain PLS, as with
PCR, we could have used penalization to derive sparse PLS components
[\citet{chun2010}], but we instead opted to build on the
aforementioned approach of \citet{johnstone2009} to sparse PCA. A
natural PLS analogue of that approach is to select those features
$\tilde{x}_j$ whose covariance with $y$ has the greatest magnitude.
This results in the sparse PLS criterion
%
\begin{equation}
\label{splscrit}\bigl\llVert \by-\bT\bdelta- \tilde{\bX }^{\dagger} \tilde{
\bR}^{\dagger}_m\bgamma\bigr\rrVert ^2;
\end{equation}
here $\tilde{\bX}^{\dagger}$ consists of the $c$ columns of $\tilde
{\bX}$ having highest covariance with $\by$, and the columns of
$\tilde{\bR}^{\dagger}_m$ are defined analogously to those of
$\tilde{\bR}_m$ in \eqref{nsplscrit}. As for PCR, the least-squares
minimizer $(\hat{\bdelta},\hat{\bgamma})$ of \eqref{splscrit}
leads directly to estimates of the wavelet coefficients $\tilde{\bbeta
}$ and of the resulting coefficient image $\bbeta$.

Our PLS algorithm is a wavelet-domain counterpart of the spline-based
functional PLS procedure denoted by FPLS$_C$ in \citet
{reiss2007}. We note that \citet{preda2005} and \citet
{delaigle2012annals} have proposed more explicitly functional
formulations of PLS, based on covariance operators on function spaces.

\subsection{Wavelet-domain elastic net}
Since wavelet bases are well suited for sparse representation of
functions, recent work has considered combining them with
sparsity-inducing penalties, both for semiparametric regression
[\citet{wand2011}] and for regression with functional or image
predictors [\citet{zhao2012,wang2014,zhao2014}]. The latter
papers focused on $\ell_1$ penalization, also known as the lasso
[\citet{tibshirani1996}], in the wavelet domain. Alternatives to
the lasso include the SCAD penalty [\citet{fan2001}] and the
adaptive lasso [\citet{zou2006adaptive}]. Here we consider
the elastic net (EN) estimator for wavelet-domain model \eqref
{wdform}, which minimizes
%
\begin{equation}
\label{encrit}\llVert \by-\bT\bdelta-\tilde{\bX}\tilde {\bbeta}\rrVert
^2+\lambda \bigl[\alpha\llVert \tilde{\bbeta}\rrVert
_1+(1-\alpha )\llVert \tilde{\bbeta}\rrVert _2^2
\bigr]
\end{equation}
over $(\bdelta,\tilde{\bbeta})$, for a regularization parameter
$\lambda>0$ and a mixing parameter $\alpha\in[0,1]$ which controls
the relative strength of the $\ell_1$ and $\ell_2$ penalties on the
coefficients [\citet{zou2005}].

In the original nomenclature of \citet{zou2005}, the minimizer of
\eqref{encrit} is the ``na\"{i}ve'' EN, whereas EN is a rescaled
version. Since we shall make use of the generalized linear extension of
EN as implemented by \citet{friedman2010}, we follow these
authors in omitting the rescaling step. When $\alpha>0$, the $\ell_1$
penalty shrinks small coefficients to zero, leading to a sparse wavelet
representation. The wavelet-domain lasso is obtained when $\alpha=1$.
As explained by \citet{zou2005}, given a group of important
features that are highly correlated, the lasso tends to select just
one, whereas EN selects the entire group, which is often
preferable---even in the wavelet domain, notwithstanding the
``whitening'' property of the discrete wavelet transform.

\subsection{Summary: Alternative routes to sparsity}\label{sarts}
All three of the above methods seek to represent the coefficient image
$\beta(\cdot)$ sparsely, as a linear combination of a subset of the
wavelet basis functions, but they deploy very different strategies to
choose that subset. The $\ell_1$ penalty in the elastic net criterion
\eqref{encrit} has the effect of shrinking small coefficients to zero.
This can be interpreted as imposing a prior that favors a sparse
estimate. The PCR criterion \eqref{spcrcrit} eliminates basis elements
\emph{before} performing regression, based on an implicit assumption
that those basis elements with low variance in the data have little to
contribute to the coefficient image. This assumption is broadly
consistent, on the one hand, with the assumption of \citet
{johnstone2009} that such basis elements are merely capturing noise;
and, on the other hand, with the underlying assumption of PCR, namely,
that the highest-variance principal components are most relevant in
regression [see \citet{cook2007} for some relevant discussion].
The PLS criterion \eqref{splscrit} likewise lets the data determine
which basis elements to include; but here, instead of considering only
the wavelet-transformed image data $\tilde{\bX}$ as in PCR, we define
relevant components by iteratively maximizing covariance with the
responses $\by$.

\subsection{Extension to the generalized linear case}\label{x2glm}
The above three wavelet-domain algorithms can be straightforwardly
extended from linear to generalized linear models (GLMs) of the form
%
\begin{equation}
\label{wdglm}g\bigl[E(\by)\bigr]=\bT\bdelta+ \tilde{\bX }\tilde{\bbeta},
\end{equation}
for a link function $g$, generalizing \eqref{wdform}. For PCR, one
simply fits a GLM, as opposed to a linear model, to the sparse PCs. For
the elastic net, the \texttt{glmnet} algorithm of \citet
{friedman2010} is available for the generalized linear case.

PLS is sometimes performed in an iteratively reweighted manner for GLMs
[\citet{marx1996}], but in high-dimensional settings, such
algorithms may require nontrivial modification [e.g., \citet
{ding2005}] to avoid convergence \mbox{problems}. Here we view PLS as a
generic approach to constructing relevant components, which may be
employed beyond the linear regression setting [e.g., \citet
{nguyen2002,delaigle2012jrssb}]. Thus, we construct PLS components
exactly as
we would for a linear model, but then use these components to fit a GLM.

\subsection{Tuning parameter selection}\label{tupa}
For wavelet-domain PCR and PLS, three tuning parameters must be
selected:\vspace*{1pt} the resolution-level parameter $j_0$; the number $c$ of
wavelet coefficients to retain [i.e., the number of columns of $\tilde
{\bX}^*$ in \eqref{spcrcrit} or of $\tilde{\bX}^{\dagger}$ in
\eqref{splscrit}]; and the number $m$ of PCs or PLS components. We
generally fix $j_0=4$, since we have found that resolution level to be
generally either optimal or near-optimal as measured by
cross-validation (CV). For wavelet-domain elastic net, one must choose
$j_0$ and the two penalty parameters $\alpha$ and $\lambda$ in \eqref
{encrit}, but we again prefer to fix $j_0=4$.

These tuning parameters are chosen by repeated $K$-fold CV. In the
$r$th of $R$ repetitions we divide the data points $(y_i,\mathbf{t}_i,\mathbf{x}_i)$ ($i=1,\ldots,n$) into $K$
equal-sized validation sets indexed by $I_{r,1},\ldots,I_{r,K}$.
We can then choose the tuning parameters to minimize the CV score
%
\begin{equation}
\label{rkcv}\frac{1}{RK}\sum_{r=1}^R
\sum_{k=1}^K \sum
_{i\in I_{r,k}}L(y_i;\hat{\bdelta}_{-r,k},\hat{
\tilde{\bbeta }}_{-r,k}),
\end{equation}
where $\hat{\bdelta}_{-r,k},\hat{\tilde{\bbeta}}_{-r,k}$ are the
estimates that result when model \eqref{wdglm} is fitted (by PCR, PLS
or EN) with the observations indexed by $I_{r,k}$ excluded, and $L$ is
an appropriate loss function. For linear regression the standard loss
function is the squared error $L(\by_i;\bdelta,\tilde{\bbeta
})=(y_i-\mathbf{t}_i^T\bdelta-\tilde{\mathbf{x}}_i^T\tilde{\bbeta})^2$. For the generalized linear case,
following \citet{zhu2004}, we use the deviance $D(y_i;\bdelta,\tilde{\bbeta})$ as the loss function. Specifically for logistic
regression, unusually large summands can dominate criterion~\eqref
{rkcv}. Therefore, similarly to \citet{chi2013}, we instead
choose the tuning parameters by a robust CV score that takes the median
rather than the mean over each set of $K$ validation sets:
%
\begin{equation}
\label{robcv}\frac{1}{R}\sum_{r=1}^R
\operatorname{median}\limits_{k\in\{1,\ldots,K\}} \sum_{i\in I_{r,k}}D(y_i;
\hat{\bdelta }_{-r,k},\hat{\tilde{\bbeta}}_{-r,k}).
\end{equation}

\section{Comparative simulation study}\label{simsec}
To test the performance of our methods with realistic image predictors,
we created a data set based on the positron emission tomography (PET)
data previously studied by \citet{reiss2010}. That data set
included axial slices from 33 amyloid beta maps, from which we
extracted a square region of $64 \times64$ voxels. To generate a
larger sample of $n=500$ images, we applied a procedure similar to that
of \citet{goldsmith2013}:
\begin{longlist}[2.]
\item[1.] We estimated the
(vectorized) principal components (eigenimages)
\[
\hat{\brho}_1, \dots, \hat{\brho}_{32}\in{\mathbb
R}^{64^2},
\]
with corresponding eigenvalues $\lambda_1, \dots, \lambda_{32}$.

\item[2.] For $i=1,\ldots,500$, we generated the $i$th simulated predictor
image as $\mathbf{x}_i=\sum_{j=1}^{32}c_{ij}\hat{\brho
}_j$, with the $c_{ij}$'s simulated independently from the $N(0,\lambda
_j)$ distribution.
\end{longlist}
In step 1 above we used the sparse PCA method of \citet
{johnstone2009}, including the 492 wavelet coefficients having the
highest variance. This number of wavelet coefficients was sufficient to
capture 99.5\% of the ``excess'' variance, in the sense of Section~4.2
of \citet{johnstone2009}.

%
\begin{figure}[t]

\includegraphics{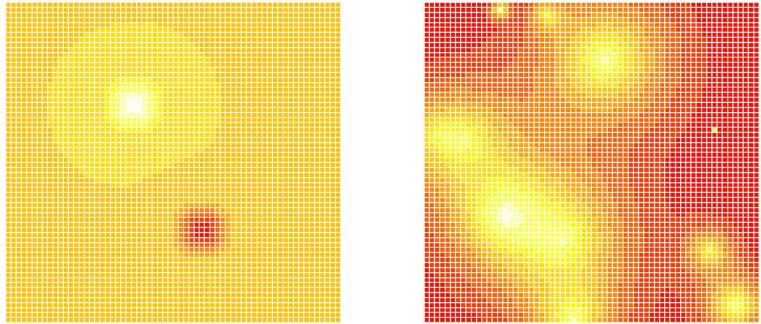}

\caption{Coefficient images $\bbeta^{(1)}$ (left) and $\bbeta^{(2)}$
(right) used in the simulation study.}\label{figomega}
\end{figure}

We used two different true coefficient images $\bbeta\in{\mathbb
R}^{64^2}$, which are shown in Figure~\ref{figomega}. The first image
$\bbeta^{(1)}$ is similar to that used by \citet{goldsmith2013}.
Taking its domain to be $[1,64]^2$, this coefficient image is given by
$\beta^{(1)}=g_1-g_2$, where $g_1,g_2$ are the densities of the
bivariate normal distributions
\[
N \left[\pmatrix{30
\cr
20}, 10\bI_2 \right]\quad\mbox{and}\quad N \left[ \pmatrix{20
\cr
55},
10\bI_2 \right],
\]
respectively. The second image $\bbeta^{(2)}$ is a two-dimensional
analogue of the ``bumps'' function used by \citet{donoho1994},
and many subsequent authors, to illustrate the properties of wavelets.
%
%

We then simulated continuous or binary outcomes $y_1,\ldots,y_n$ with
specified approximate values of the coefficient of determination $R^2$,
in the sense detailed in Supplementary Appendix~A.1
[\citet{reiss2015}]. We generated 100 sets of $n=500$ continuous
outcomes and 100 sets of 500 binary outcomes, for each of the $R^2$
values $0.1, 0.5$.

We compared the performance of the three wavelet-domain methods
described in Section~\ref{twda} with three analogous ``voxel-domain''
methods, that is, sparse PCR, sparse PLS and elastic net without
transformation to the wavelet domain. The wavelet- and voxel-domain
methods are denoted by WPCR, WPLS and WNet and by VPCR, VPLS and VNet,
respectively. We also included the $B$-spline-based functional PCR
method (``FPCR$_R$,'' or simply FPCR) of \citeauthor{reiss2007} (\citeyear{reiss2007,reiss2010}). Tuning parameter selection was as described in
Supplementary Appendix~A.1 [\citet{reiss2015}].

Performance was evaluated in terms of estimation error and prediction
error. Estimation error is defined by the scaled mean squared error
(MSE) $\llVert  \hat{\bbeta} - \bbeta\rrVert  ^2/\llVert  \bbeta\rrVert  ^2$, where $\bbeta,\hat{\bbeta}$ are the true and estimated coefficient images.
Prediction error is defined using a separate set of outcomes
$y_1^*,\ldots,y_n^*$, generated from the same conditional distribution
as $y_1,\ldots,y_n$. We use the scaled mean squared prediction error
$\frac{1}{n\sigma^2}\sum_{i=1}^n(y_i^*-\hat{y}_i)^2$ as our\vspace*{1pt}
criterion for linear regression and the mean of the deviances of
$y_1^*,\ldots,y_n^*$ for logistic regression.

%
\begin{figure}[b]

\includegraphics{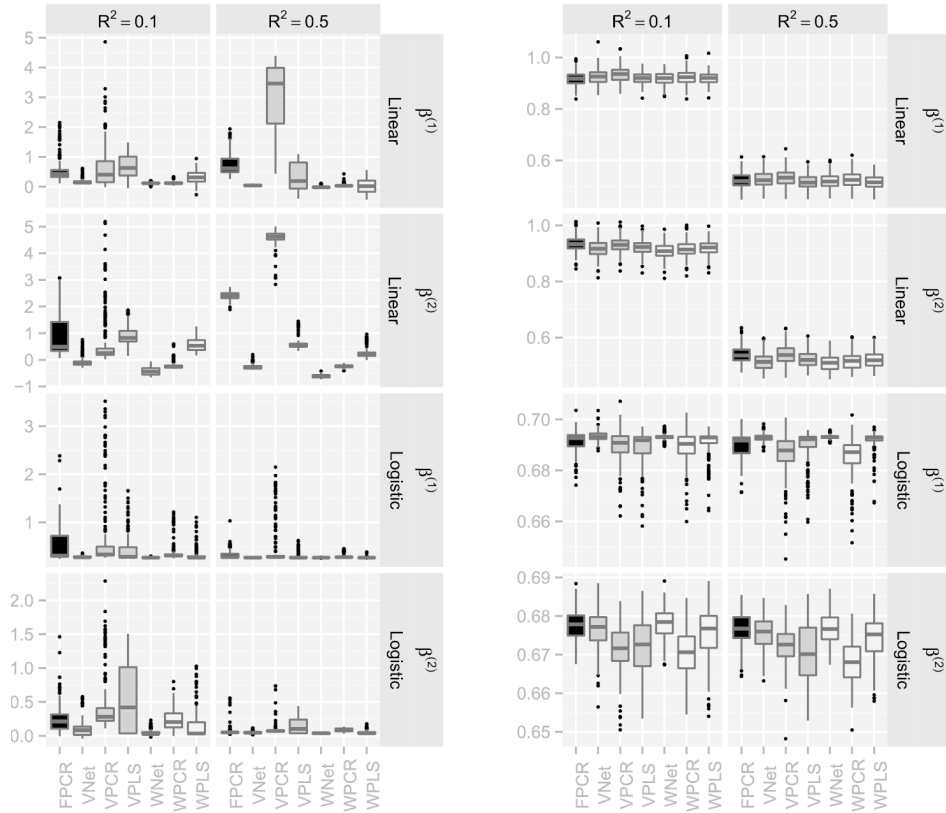}

\caption{Estimation error, displayed as $\operatorname{log}(\mbox{scaled MSE})$ (left
subfigure), and prediction error (right subfigure) in the simulation
study.}\label{simbox}
\end{figure}

Figure~\ref{simbox} presents boxplots of the results. In general, all
seven methods differ only slightly in prediction error. Much greater
differences are seen for estimation error. Compared with the
corresponding voxel-domain methods, the estimation MSE for wavelet
methods is either roughly equal or clearly lower on average, and the
variability of the MSE is often much lower. The wavelet methods also
markedly outperform $B$-spline-based FPCR. Somewhat contrary to
expectation, the superior performance of wavelet methods is not clearly
more pronounced for $\bbeta^{(2)}$ than for~$\bbeta^{(1)}$.

\begin{figure}

\includegraphics{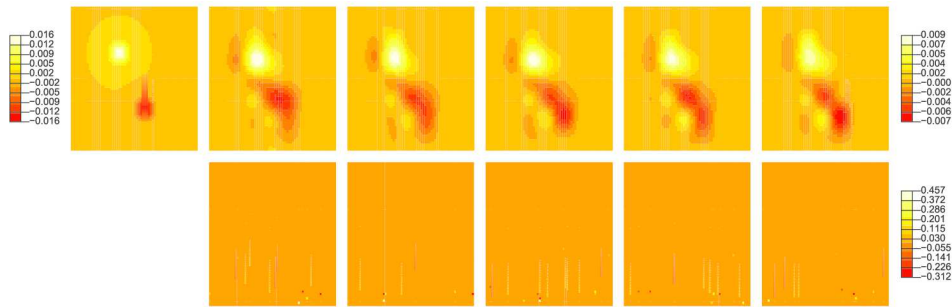}

\caption{True coefficient function $\bbeta^{(1)}$ from the
comparative simulation study (top left) compared with five training-set
coefficient function estimates (for data simulated under $R^2=1$
setting) based on wavelet-domain lasso (other top panels) and
voxel-domain lasso (bottom panels). The wavelet-based estimates are
reasonably accurate, while each of the voxel-domain estimates has about
20--25 scattered voxels with nonzero values. Note the unequal
scales.}\label{show5}
\end{figure}

While the wavelet-domain methods do not clearly attain lower estimation
error than voxel-domain methods for logistic regression with $R^2=0.5$,
they do appear superior for the $R^2=0.1$ setting (which seems more
realistic) and for linear regression. Moreover, qualitatively,
wavelet-domain modeling helps to capture the main features of the
coefficient image. Figure~\ref{show5} displays an example of the
training-set estimates derived by wavelet-domain lasso versus ordinary
lasso. The wavelet-domain estimates are clearly more similar to each
other and to the true coefficient image than are the ordinary lasso estimates.\looseness=-1
%

The wavelet-domain EN appears to have a slight edge overall compared
with PCR and PLS. For this reason, and because wavelet EN (or at least
its special case, the lasso) are now somewhat established in the
literature [\citet{zhao2012,wang2014,zhao2014}], the simulations
and real-data analyses in the next two sections consider only
wavelet-domain EN.

\section{Inferential issues}\label{infsec}
We now turn to what the \hyperref[sec1]{Introduction} referred to as limitation (ii) of
predictive analyses in neuroimaging: the need for methodology to assess
the predictive value of image data, in particular, when scalar
covariates are present.

\subsection{Permutation testing}\label{statsig}
Consider testing the null hypothesis $\beta(\bs)\equiv0$ in the
general model \eqref{efspec}, \eqref{gmuspec}, that is, testing the
null parametric model $g(\mu_i)=\mathbf{t}_i^T\bdelta$
versus the alternative \eqref{gmuspec}. Informally, we are asking
whether the images have predictive value beyond the information
contained in the\vadjust{\goodbreak} scalar predictors. We propose a permutation test
procedure in which the CV criterion \eqref{rkcv} or \eqref{robcv}
serves as the test statistic. If the true-data CV falls in the left
tail of the distribution of permuted-data CV values, significance is declared.
Permutation techniques of this kind have previously appeared in the
neuroimaging and machine learning literature [\citet
{golland2003,ojala2010}].

The way the permutation distribution is constructed depends on the null
model under consideration. When $\mathbf{t}_i\equiv1$ in
\eqref{gmuspec} (no scalar covariates), one can simply permute the
responses: that is, we repeatedly reorder the responses as $y_{\pi
(1)},\ldots,y_{\pi(n)}$ for some permutation $\pi$, refit the model,
and record the CV value. For the linear model \eqref{FLM} with scalar
covariates, a common approach is to permute the residuals from the null
parametric model: that is, model \eqref{FLM} is refitted repeatedly
with the $i$th response of the form $\hat{y}_i+\hat{\varepsilon
}_{\pi(i)}$, where the hats refer to fitted values and residuals from
the model $y_i =\mathbf{t}_i^T\bdelta+\varepsilon_i$. For
some GLMs, however, such pseudo-responses based on permuted residuals
are not of the correct form (e.g., for logistic regression, they are
not binary). One can instead form pseudo-\emph{predictors}, by
regressing the predictor of interest on the nuisance covariates and
permuting the residuals from this fit.
In other words, we replace the design matrix $(\bT|\bX)$ with
%
\begin{equation}
\label{idm} \bigl[\bT | \bP_T\bX +\bPi(\bI-\bP_T)\bX
\bigr],
\end{equation}
where $\bP_T=\bT(\bT^T\bT)^{-1}\bT^T$ and $\bPi$ is a permutation
matrix. Although a similar idea was proposed by \citet
{potter2005} for (ordinary) logistic regression, we have adopted it as
our preferred permutation approach even for the linear case; see
Supplementary Appendix~B [\citet{reiss2015}] for
further discussion.

\begin{figure}

\includegraphics{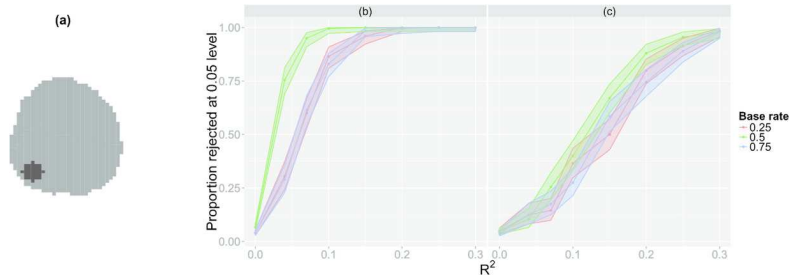}

\caption{\textup{(a)} True coefficient image $\bbeta$ used in the power study:
gray denotes 0, black denotes 1. \textup{(b)}~Estimated probability of rejecting
the null hypothesis $\bbeta=0$ as a function of $R^2$, with 95\%
confidence intervals, for model \protect\eqref{pseq}. \textup{(c)} Same, for
model \protect\eqref{pseq2}.}\label{psim}
\end{figure}

We conducted a simulation study, using the ADHD-200 image data analyzed
in Section~\ref{appsec}, to assess the type-I error rate and power of
the permutation test procedure. Here we focus on logistic regression
(see Supplementary Appendix~C [\citet{reiss2015}], for linear regression results) and the
wavelet-domain lasso. We first considered the case without scalar
covariates and generated binary responses $y_i\sim\operatorname{Bernoulli}(p_i)$, $i=1,\ldots,n=333$, where
%
\begin{equation}
\label{pseq}\log\frac{p_i}{1-p_i}=\delta_0+\mathbf{x}_i^T
\bbeta,
\end{equation}
where $\delta_0$ is a constant used to adjust the base rate
(probability of event); $\mathbf{x}_i\in{\mathbb R}^{64^2}$
is the $i$th image (expressed as a mean-zero vector); $\bbeta$ is the
true coefficient image shown in Figure~\ref{psim}(a) (similarly
vectorized), multiplied by an appropriate constant to attain a
specified value of $R^2$
(see Supplementary Appendix~A [\citet{reiss2015}], regarding the definition of $R^2$). For each
of the base rates 0.25, 0.5, 0.75 and each of the $R^2$ values 0.04, 0.07, 0.1, 0.15, 0.2, 0.25, 0.3, we simulated 200 response vectors to assess
power to reject $H_0\dvtx \bbeta=\bzero$ at the $p=0.05$ level, as well as
1000 response vectors with $\bbeta=\bzero$ ($R^2=0$) to assess the
type-I error rate.
Next we considered testing the same\vadjust{\goodbreak} null hypothesis for the model
%
\begin{equation}
\label{pseq2}\log\frac{p_i}{1-p_i}=\delta _0+t_i
\delta_1+\mathbf{x}_i^T\bbeta,
\end{equation}
with a scalar covariate $t_i$ such that $R^2$ for the submodel
$E(y_i|t_i)=t_i\delta$ is approximately 0.2. We generated the same
number of response vectors as above for each of the above $R^2$ values,
but here $R^2$ refers to the partial $R^2$ adjusting for $t_i$
(see Supplementary Appendix~A.2 [\citet{reiss2015}]).

The results, displayed in Figure~\ref{psim}(b) and (c), indicate that
the nominal \mbox{type-I} error rate is approximately attained for both
models. For a given $R^2>0$, the power is somewhat higher for model
\eqref{pseq} than for model \eqref{pseq2}, and highest for either
model when the base rate is 0.5. Evidently, for base rates closer to 0
or 1, the CV deviance under the null hypothesis tends to be lower, and
thus a stronger signal is needed to reject the null.

%

Basing a test of the hypothesis $\beta(\cdot)\equiv0$ on the
prediction performance of an estimation algorithm, rather than on an
estimate of $\beta$, is admittedly somewhat unconventional. In
neuroimaging specifically, inference typically proceeds by fitting
separate models at each voxel, and then applying some form of multiple
testing correction [\citet{nichols2012}]. In the present setting
of a single model that uses the entire image to predict a scalar
response, it might be possible to assign \mbox{$p$-}values to individual
voxels as in \citet{meinshausen2009}. In practice, however,
predictive algorithms tend to produce rather unstable estimates, as a
number of authors have acknowledged [e.g., \citet{craddock2009,honorio2012,sabuncu2012}]. Our hypothesis testing approach thus sets
the more modest inferential goal of verifying that the coefficient
image as a whole yields better-than-chance prediction.\looseness=-1

\subsection{Confounding}\label{seccfd}
For ordinary, as opposed to functional, regression, confounding is said
to occur when (i) $x$ appears predictive of $y$, but this relationship
can be attributed to a third variable $t$ such that (ii) $t$ is
predictive of $y$ and (iii) $t$ is correlated with $x$. For example,
birth order ($x$) is associated with the occurrence of Down syndrome
($y$), but this is due to the effect of the confounding variable
maternal age ($t$) [\citet{rothman2012}].

To extend the above definition to the case of a functional predictor
$x(\cdot)$, suppose that (i) $x(\cdot)$ is ostensibly related to $y$,
in the sense that $\beta(\cdot)$ is not identically zero when model
\eqref{gmuspec} includes no scalar covariates, but (ii) the scalar
variable $t$ is also predictive of $y$. A functional-predictor analogue
of point (iii) is to suppose that $t$ is correlated with $\int x(s)\hat
{\beta}(s)\,ds$, where $\hat{\beta}(\cdot)$ is an estimate obtained
with $t$ excluded from model \eqref{gmuspec}. Aside from this
``global'' analogue of (iii), it may be useful to consider a ``local''
analogue which holds if $t$ is correlated with $x(s)$, specifically for
$s$ such that $\beta(s)\neq0$; but this is somewhat less
straightforward to assess.

\section{Application: fALFF and ADHD}\label{appsec}

\subsection{ADHD-200 data set and candidate models}
We now apply the wa\-velet-domain elastic net to ``predicting'' ADHD
diagnosis using maps of fractional amplitude of low-frequency
fluctuations (fALFF) [\citet{zou2008}] from a portion of the
ADHD-200 sample referred to in the \hyperref[sec1]{Introduction} (\surl{http://fcon\_1000.projects.nitrc.org/indi/adhd200/}).
fALFF is defined as the ratio of BOLD signal power spectrum within the
0.01--0.08~Hz range to total over the entire range. \citet
{yang2011} reported altered levels of fALFF in a sample of children
with ADHD relative to controls, specifically in frontal regions. That
study relied on the traditional analytic approach in neuroimaging,
which regresses the imaged quantity (in this case fALFF) on diagnostic
group, separately at each voxel. Here we employed scalar-on-image
logistic regression, which reverses the roles of response and
predictor, to regress diagnostic group on fALFF images.
Our sample consisted of 333 individuals: 257 typically developing
controls and 76 with combined-type ADHD. The sample included 198 males
and 135 females, with age range 7--20 (see
Supplementary Appendix~D [\citet{reiss2015}], for further details).
We chose the 2D slice for which the mean across voxels of the SD of
fALFF was highest. This was the axial slice located at
$z=26$ (just dorsal to the corpus callosum) in the coordinate space of
the Montreal Neurological Institute's MNI152
template (4~mm resolution).
We fitted two models. The first was
%
\begin{equation}
\label{iom}
\operatorname{logit~Pr}(i\mbox{th subject has ADHD})=\delta+ \int
_{\mathcal S}x_i(\bs)\beta(\bs)\,d\bs,
\end{equation}
where $x_i(s)$ denotes the $i$th subject's fALFF image. The second
model was
%
\begin{equation}
\label{iom2}
\operatorname{logit~Pr}(i\mbox{th subject has ADHD})=\mathbf{t}_i^T
\bdelta+ \int_{\mathcal S}x_i(\bs )\beta(\bs)\,d\bs,
\end{equation}
where the vector $\mathbf{t}_i$ includes the $i$th subject's
age, sex, IQ and mean FD, as well as a leading 1 for the intercept.\vadjust{\goodbreak}

\begin{figure}

\includegraphics{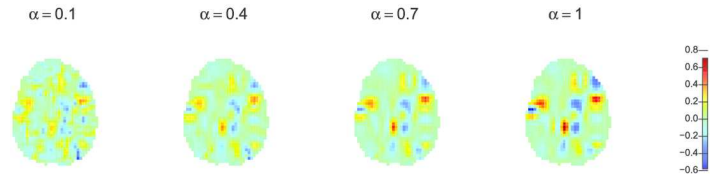}

\caption{Coefficient image estimates for model \protect\eqref{iom}
applied to the ADHD-200 data, using wavelet-domain elastic net with
four different values of the mixing parameter $\alpha$.}\vspace*{-3pt}\label{wnetfhat}
\end{figure}

\begin{figure}[b]

\includegraphics{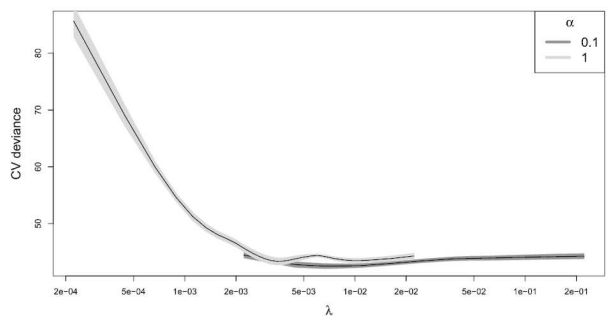}

\caption{Cross-validated deviance $+$/$-$ one approximate standard
error, for the wavelet-domain elastic net models with $\alpha
=0.1,1$.}\vspace*{-3pt}\label{wnetcv}
\end{figure}

Figure~\ref{wnetfhat} shows the coefficient images attained for model
\eqref{iom} with each value of the mixing parameter $\alpha$. As
expected, increasing values of $\alpha$ lead to more-sparse estimates
in the wavelet domain, and hence in the voxel domain. Figure~\ref
{wnetcv} shows the CV deviance as a function of $\lambda$ for $\alpha
=0.1$, which had the lowest CV deviance overall, as well as for $\alpha=1$.

%

%

\begin{figure}

\includegraphics{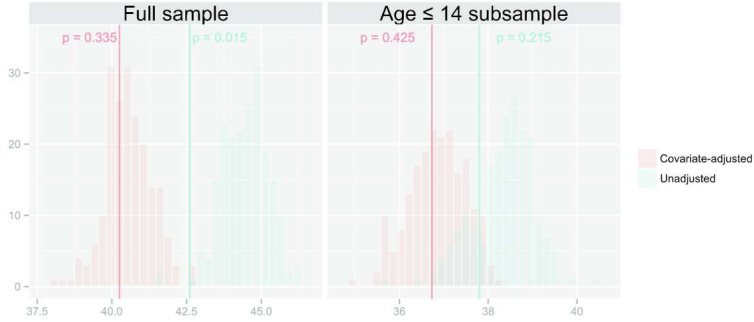}

\caption{Permutation test results. For the full sample (left), a
significant effect of the fALFF images is seen in model \protect\eqref
{iom}, but not in model \protect\eqref{iom2}, which adjusts for
scalar covariates. When only younger individuals are included (right),
neither model shows a significant fALFF effect.}
\label{ptr}
\end{figure}

The left subfigure of Figure~\ref{ptr} shows that the CV deviance lies
in the left tail of the permutation distribution for model \eqref
{iom}, indicating a significant effect of the fALFF image predictors
($p=0.015$). However, with the scalar covariate adjustment of model
\eqref{iom2}, this effect disappears. The next subsection examines
more closely how the scalar covariates may be acting as confounders.
%

Our test of model \eqref{iom} entailed 999 permuted-data fits with
four candidate values of $\alpha$ and 100 of $\lambda$, requiring
14.25 hours on an Intel Xeon E5-2670 processor running at 2.6 GHz. In
practice, we recommend parallelizing the permutations via cluster
computing to make the computation time more manageable. In addition,
truncated sequential probability ratio tests [\citet{fay2007}]
could in some cases reduce computation time via early stopping. We also
explored fitting model \eqref{iom} with the full 3D fALFF images as
predictors; see Supplementary Appendix~E [\citet
{reiss2015}].

\subsection{Assessing and remedying confounding}
\begin{figure}[b]

\includegraphics{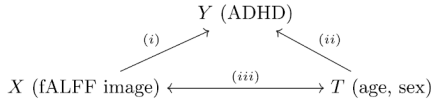}

\caption{Relationships among a putative predictor $X$, outcome $Y$ and
confounder $T$ (see Section~\protect\ref{seccfd}), illustrated with
respect to the ADHD-200 data.}\label{cfd-diag}
\end{figure}

\begin{figure}[b]

\includegraphics{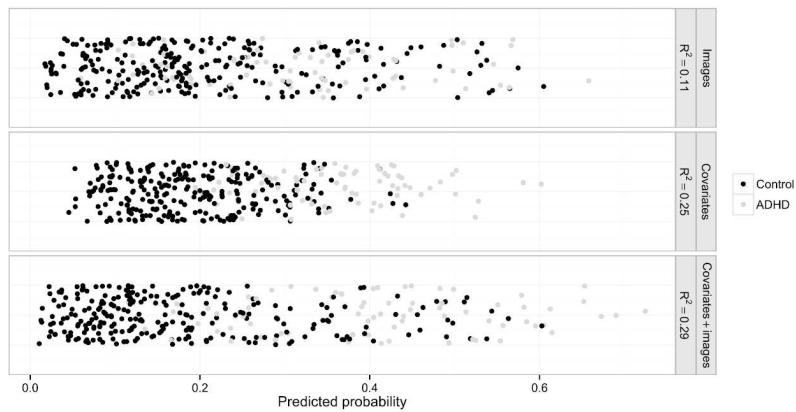}

\caption{Predicted probabilities of ADHD diagnosis, according to the
images-only model \protect\eqref{iom}; an ordinary logistic
regression with the four scalar covariates; and model \protect\eqref
{iom2}, which includes both. Also shown are the $R^2$ values, as
defined in Supplementary Appendix~\textup{A.1}, for the
three models.}
\label{sep}
\end{figure}

As discussed in Section~\ref{seccfd}, the notion of confounding
entails three elements (see Figure~\ref{cfd-diag}). Point (i), an\vadjust{\goodbreak}
apparent effect of the image predictor fALFF on diagnosis, was
established by the above permutation test result for model \eqref{iom}.
To check point (ii) of the definition for each of the four scalar
covariates under consideration, we performed an ordinary logistic
regression with diagnosis (1${}={}$ADHD, 0${}={}$control) as response and the above
four scalar predictors. In Table~\ref{plr} (at left), sex, age and IQ
are all seen to be significantly related to diagnosis. See also
Figure~\ref{sep}, which compares the fitted probabilities from this
ordinary logistic regression with those resulting from models \eqref{iom} and~\eqref{iom2}. The scalar-covariates model is seen to
separate the two groups (black vs. gray dots) quite well; the image
predictors increase the spread of the predicted probabilities without
clearly improving the two groups' separation. Based on these results,
each of these three variables may be acting as a confounder.
%
%
%
\begin{table}
\tabcolsep=0pt
\caption{To examine element \textup{(ii)} of confounding, an ordinary logistic
regression was fitted with the four scalar predictors and with ADHD
diagnosis as response; the resulting estimates are shown with 95\%
confidence intervals. For element \textup{(iii)}, we display the correlations of
each predictor  with the logit probabilities estimated by fitting model~\protect\eqref{iom}}\label{plr}
\begin{tabular*}{\tablewidth}{@{\extracolsep{\fill}}@{}ld{2.15}d{1.5}d{2.15}c@{}}
\hline
&\multicolumn{2}{c}{\textbf{(ii)}} &\multicolumn{2}{c@{}}{\textbf{(iii)}}\\[-6pt]
&\multicolumn{2}{c}{\hrulefill} &\multicolumn{2}{c@{}}{\hrulefill}\\
& \multicolumn{1}{c}{\textbf{Log odds ratio}}
& \multicolumn{1}{c}{\textbf{$\bolds{p}$-value}}
& \multicolumn{1}{c}{\textbf{Correlation}}
& \multicolumn{1}{c@{}}{\textbf{$\bolds{p}$-value}}\\
\hline
Intercept & 3.90~(1.11, 6.78) & 0.007&& \\
Sex~(M--F) & 1.26~(0.65, 1.91) & 0.00008 &0.14~(0.03,0.24)&0.011\\
Age & -0.20~(-0.32, -0.09) & 0.0005&-0.35~(-0.44,-0.25)& $6\times
10^{-11}$\\
IQ & -0.03~(-0.05, -0.01) & 0.003 &-0.09~(-0.19,0.02)&0.10\\
Mean FD & -2.51~(-8.80, 3.56) & 0.42&-0.04~(-0.15,0.07)&0.47 \\
\hline
\end{tabular*}
\end{table}
%
%

Next we consider point (iii), that is, the correlations of each scalar
covariate with $\int_{\mathcal S}x_i(\bs)\hat{\beta}(\bs)\,d\bs$,
where $\hat{\beta}$ is the coefficient image estimate from the
fALFF-only model \eqref{iom} or, equivalently, with the predicted
logit probability of ADHD from that model. The results, shown at right
in Table~\ref{plr}, point to age and sex as the principal confounders.
(Here sex was treated as a binary variable, with 1 for male and 0 for
female; a $t$-test and a Mann--Whitney test yielded similar results.)
``Local'' examination in the sense of Section~\ref{seccfd} reveals
that the fALFF $x(\bs)$ tends to be higher in males and in younger
individuals for many voxels $\bs$; and such regions overlap
considerably with those in which $\hat{\beta}(\bs)>0$. In other
words, the ostensible association between fALFF and ADHD likely
reflects the dependence of fALFF on age and sex, which in turn are
related to ADHD in our sample.

Further inspection revealed that, of the 67 individuals with age above
14.0, only 8 had ADHD, with maximum age 17.43---whereas the controls
had ages as high as 20.45. This led us to suspect that these older
individuals might be driving the confounding with age that results in a
spurious effect of fALFF on diagnosis. To investigate this possibility,
we repeated the analysis using only the 266 individuals of age 14.0 or lower.
Figure~\ref{ptr} shows that in this subsample, the fALFF effect is no
longer significant, even without adjusting for the scalar covariates.
Moreover, given how far the test statistic is from the left tail of the
permutation distribution, it seems unlikely that the loss of
significance is due merely to the lower sample size.

In general, absent careful matching at the design stage, it would be
advisable to match the two diagnostic groups optimally on a complete
set of clearly relevant variables, via algorithms such as those
described in \citet{rosenbaum2010}. Our aim here, however, was to
show how a straightforward new notion of confounding for functional
predictors can be used to identify a principal scalar confounder, whose
impact can be removed by the crude device of simply truncating the age range.

\section{Discussion}\label{discuss}
Our analysis in Section~\ref{appsec} included only one imaging
modality and only a subset of the individuals from the ADHD-200 Global
Competition database. At any rate, our essentially negative result is
consistent with the finding [\citet{brown2012}] that diagnostic
accuracy was optimized by basing prediction on scalar predictors, while
ignoring the image data.
In a blog comment on that outcome, cited both by \citet{adhd2012}
and by \citet{brown2012}, the neuroscientist Russ Poldrack
suggested that ``any successful imaging-based decoding could have been
relying upon correlates of those variables rather than truly decoding a
correlate of the disease.'' Stated a bit differently, the competing
teams' successes in using the image data to predict diagnosis may have
been brought about by confounding. But there appear to have been few
attempts, if any, to study systematically how confounding may give rise
to spurious relationships between quantitative image data and clinical
variables. Similarly, analyses of the ADHD-200 data, and related work
on brain ``decoding,'' have devoted little attention to formally
testing the contribution of imaging data to prediction of scalar responses
[but see \citet{Rei15}].

As we have shown, these two interrelated issues---testing the effect of
image predictors and investigating possible confounders---can be
handled straightforwardly within our scalar-on-image regression
framework. The permutation test procedure of Section~\ref{statsig}
found a statistically significant relationship between fALFF images and
ADHD diagnosis, but this disappeared when four scalar covariates were
adjusted for. Further examination, in light of our extension of the
notion of confounding to functional/image predictors in Section~\ref
{seccfd}, pointed to age and sex as the key confounders.

The ADHD-200 project is one of a number of recent initiatives to make
large samples of neuroimaging data publicly available [\citet
{milham2012}]. These initiatives have been a boon for statistical
methodology development, but it must be borne in mind that even as
neuroimaging sample sizes increase rapidly, they remain much smaller
than the data dimension. No approach to scalar-on-image regression can
completely escape the ensuing nonidentifi\-a\-bil\-ity of the
coefficient image. We can, however, (i) put forth assumptions, likely
to hold approximately in practice, that reduce the effective dimension
of the coefficient image; and (ii) employ multiple methods in the hope
that these will converge upon similar coefficient image estimates, at
least when the signal is sufficiently strong.

With these considerations in mind, we have introduced three methods for
scalar-on-image regression, each relying on a different set of
assumptions to achieve \mbox{dimension} reduction in the wavelet domain.
Implementations of these three methods, for 2D and 3D image data, are
provided in the \texttt{refund.wave} package [\citet{refundwave}]
for R [\citet{R}], available at \surl{http://cran.r-project.org/web/packages/\\refund.wave}. This new package,
a spinoff of the \texttt{refund} package [\citet{refund}],
relies on the \texttt{wavethresh} package [\citet{nason2013}]
for wavelet decomposition and reconstruction.

As discussed in Section~\ref{metalg}, the three methods described here
are merely three instances of a meta-algorithm for scalar-on-image
regression. The\break \texttt{refund.wave} package allows for
straightforward incorporation of alternative penalties, and other
extensions may allow for more refined wavelet-domain algorithms, which
may improve the stability and reproducibility of the coefficient image
estimates [\citet{rasmussen2012}]. For instance, in wavelet-based
nonparametric regression, thresholding is often performed in a
level-specific manner. Analogously, it might be appropriate to modify
criterion \eqref{encrit} so as to differentially penalize coefficients
at different levels. One might also employ resampling techniques [cf.
\citet{meinshausen2010}] to select those wavelet basis elements
that are consistently predictive of the outcome. Finally, wavelets
whose domain is anatomically customized, such as the wavelets defined
on the cortex by \citet{ozkaya2011}, offer a promising new way to
confine the analysis to relevant portions of the brain.\looseness=-1


\section*{Acknowledgments}
The authors are grateful to the Editor, Karen Kafa\-dar, and to the
Associate Editor and referees, whose feedback led to major improvements
in the paper; to Adam Ciarleglio, for contributions to software
implementation; to Xavier Castellanos, Samuele Cortese, Cameron
Craddock, Brett Lullo, Eva Petkova, Fabian Scheipl and Victor Solo for
helpful discussions about our methodology and its application; to Jeff
Goldsmith, Lei Huang and Ciprian Crainiceanu for sharing their insights
as well as a preprint of \citet{goldsmith2013}; and to the
ADHD-200 Consortium (\surl{http://fcon\_1000.projects.nitrc.org/indi/adhd200/}) and the Neuro Bureau
(\surl{http://neurobureau.projects.nitrc.org/}) for making the fMRI data set
publicly available. In addition to the funding sources listed on the
first page, the first author thanks the National Science Foundation for
its support of the Statistical and Applied Mathematical Sciences
Institute, whose Summer 2013 Program on Neuroimaging Data Analysis
provided a valuable opportunity to present part of this research. This
work utilized computing resources at the High Performance Computing
Facility of the Center for Health Informatics and Bioinformatics at New
York University Langone Medical Center.

\begin{supplement}
\stitle{Supplementary appendices}
\slink[doi]{10.1214/15-AOAS829SUPP} 
\sdatatype{.pdf}
\sfilename{aoas829\_supp.pdf}
\sdescription{Description of simulation details, permutation of
residuals for the proposed test procedure, a power study, selection of
a subsample from the ADHD-200 data set, and results with 3D predictors.}
\end{supplement}

%

\printaddresses
\end{document}